\documentclass{llncs}
\pdfoutput=1
\usepackage{graphicx}
\usepackage{multicol}
\usepackage{multirow}
\usepackage{amsmath}
\usepackage{url}

\begin{document}

\title{Nesting Probabilistic Inference}
\author{Theofrastos~Mantadelis and Gerda~Janssens}
\institute{Department of Computer Science, Katholieke Universiteit Leuven\\
           \url{{firstname.lastname}@cs.kuleuven.be}}

\maketitle

\begin{abstract}
When doing inference in ProbLog, a probabilistic extension of Prolog, we extend SLD resolution with some additional bookkeeping. This additional information is used to compute the probabilistic results for a probabilistic query. In Prolog's SLD, goals are nested very naturally. In ProbLog's SLD, nesting probabilistic queries interferes with the probabilistic bookkeeping. In order to support nested probabilistic inference we propose the notion of a parametrised ProbLog engine. Nesting becomes possible by suspending and resuming instances of ProbLog engines. With our approach we realise several extensions of ProbLog such as meta-calls, negation, and answers of probabilistic goals.
\end{abstract}

\section{Introduction}
In Prolog, we typically write a program and then we formulate queries in terms of the program predicates. The program predicates can call new Prolog queries, affectively nesting them. A Prolog system uses SLD resolution to compute the failure or success of the queries and in case of success the answer substitutions. During SLD resolution queries are nested in a very natural way: in order for a particular query to be proved the conjunction of nested queries must be proved. Results of queries are always used in the same way: failure causes backtracking and on success we use the answer substitution. Meta-programming makes it possible to construct some of the nested queries at run-time, but once the Prolog system has mapped such a callable term to a query, execution continues as if the query was known at compilation time.

ProbLog~\cite{KIMMIG08} is a probabilistic extension of Prolog: facts can be labelled with probabilities. Labelling e.g. an \verb|edge/2| fact with a probability $p$ indicates that the edge exists with probability $p$. The basic query in ProbLog is the computation of the success probability $(Result)$ of a query $(Query)$ for a given ProbLog program: \verb|problog_inference(Inference, Query, Result)|. As the ProbLog system supports several inference methods, we also indicate which inference method $(Inference)$ will be used. In order to compute the probability, the inference methods need to do some bookkeeping about the probabilistic facts used for proving the query. Current probabilistic logic programming languages such as PHA~\cite{POOLE93}, PRISM~\cite{SATO01}, PITA~\cite{RigSwi10-ICLP10-IC}, do not support nesting of inference. We are presenting an approach for nesting inference. Nesting inference in a probabilistic logic programming is required for implementing high order probabilistic calls. Nesting inference in ProbLog interferes with the necessary bookkeeping performed. Moreover, ProbLog queries can return different kinds of results. For a ground query, ProbLog can compute its success probability. For a non-ground query, ProbLog can compute the different answers for the query together with their respective probabilities.

Instead of computing the probability, ProbLog can also return detailed information about the annotated facts used during the proofs of the query, for example the corresponding DNF. Therefore, a more general ProbLog query has the form \verb|problog_inference(Inference, Query, ResultType, Result)|. All three kinds of results have their uses as we illustrate in this paper.

The contribution of this paper is a general solution for nesting ProbLog queries, performing nested inference. Our solution allows to suspend the inference of a query $q_c$ together with its relevant probabilistic information, to start the inference of a new query $q_n$ and to compute its desired result, and then use the result of $q_n$ when resuming the inference of $q_c$. Our approach is based on ProbLog engines. A ProbLog engine has a set of parameters and a state. Every different instantiation of these parameters implement a different ProbLog inference method. By suspending the execution of the current ProbLog engine and creating a new one, we are able to support nested inference. To suspended and resume ProbLog engines we use a stack.

We introduce ProbLog in Section~\ref{sec:problog} and describe some common inference methods. In Section~\ref{sec:motivation} we present the limitation of the existing system and motivate the need for probabilistic meta-calls. Then we propose a way to overcome this limitations by the usage of ProbLog engines in Section~\ref{sec:problog_engine}. Then follows some new primitives and some examples of their usage in Section~\ref{sec:meta-calls}. The experiments are in Section~\ref{sec:experiments} and finally, Section~\ref{sec:conclusions} concludes.

\section{ProbLog}
\label{sec:problog}
ProbLog is a probabilistic extension of Prolog inspired by typical machine learning applications. It is developed as a simple but powerful probabilistic logic programming language, and can be used e.g. for mining large biological networks (where nodes represent genes, proteins, and so on), with probability labels on their edges.

\begin{figure}
  \begin{tabular}{cl}
\multirow{7}{*}{\includegraphics[scale=0.40]{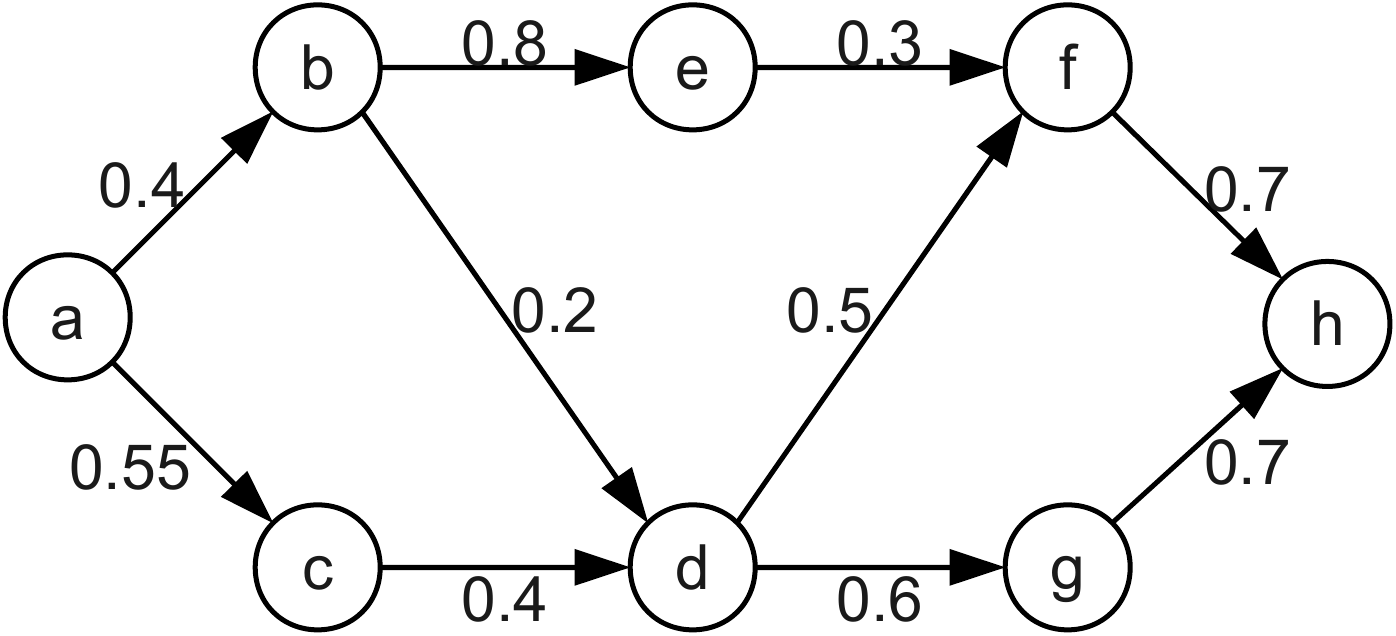}} &\verb|0.40 :: edge(a,b).  0.55 :: edge(a,c).|\\
&\verb|0.80 :: edge(b,e).  0.20 :: edge(b,d).|\\
&\verb|0.40 :: edge(c,d).  0.30 :: edge(e,f).|\\
&\verb|0.50 :: edge(d,f).  0.60 :: edge(d,g).|\\
&\verb|0.70 :: edge(f,h).  0.70 :: edge(g,h).|\\
&\verb|path(X, Y) :- edge(X, Y).|\\
&\verb|path(X, Y) :- edge(X, Z), path(Z, Y).|\\
&\\
\textbf{(a)} Probabilistic graph & \multicolumn{1}{c}{\textbf{(b)} ProbLog program}
  \end{tabular}

  \caption{An example of a probabilistic graph and the corresponding ProbLog program.}
  \label{fig:problog_example}
\end{figure}

As illustrated in Figure~\ref{fig:problog_example}, the syntax of a ProbLog program $T$ is similar to that of a Prolog program: it consists of facts and relations between them, but in the case of ProbLog a label is attached to some of the facts. That is, the program can be split into a set of labelled facts, where each $p_i :: f_i$  defines a fact $f_i$ with probability of occurrence $p_i$, and a Prolog program using those facts, which encodes the background knowledge ($BK$). We denote the set of all $f_i$ (without probability label) by $L_T$. Probabilistic facts correspond to mutually independent random variables (RVs), which together define a probability distribution over all ground logic programs $L \subseteq L_T$:
\begin{equation}
    P(L|T)=\prod\nolimits_{f_i\in L}p_i\prod\nolimits_{f_i\in L_T\backslash L}(1-p_i)\ldotp
\label{eq:subprog}
\end{equation}
We use the term \emph{possible world} to denote the least Herbrand model of such a subprogram $L$ together with the background knowledge $BK$ and, by slight abuse of notation, use $L$ to refer to both the set of sampled facts and the corresponding world.

Figure~\ref{fig:problog_example} shows a typical example of a probabilistic graph encoded in ProbLog. One can query the probability that a path exists between two nodes in the graph. As it can be noticed from the graph of Figure~\ref{fig:problog_example}, there are several possible paths between two nodes. For example between nodes $b$ and $f$, we have two possible paths: $b \rightarrow e \rightarrow f$ and $b \rightarrow d \rightarrow f$. In ProbLog, querying for the probability of \verb|path(b, f)| means asking for the probability that a randomly selected subgraph contains a path from $b$ to $f$. Such subgraphs can contain the edges of the path $b \rightarrow e \rightarrow f$ or those of the path $b \rightarrow d \rightarrow f$, but also all of them or even many more. The success probability $P_s(q|T)$ of a query $q$ can now be defined as follows:
\begin{equation}
  P_s(q|T) = \sum_{L \subseteq L_T}{P(q|L) \cdot P(L|T)}
\label{eq:success}
\end{equation}
where $P(q|L)$ is $1$ if there is a substitution $\theta$ such that $q\theta$ is entailed by the union of $L$ and the background knowledge $(L \cup BK \models q\theta)$, and $0$ otherwise. Equation~(\ref{eq:success}) states that the success probability of the query \verb|path(b, f)| can be calculated by summing the probabilities of all subgraphs which include at least one path connecting nodes $b$ and $f$.

For our example, the success probability as given in Equation (\ref{eq:success}) is easily computed even by hand: the success probability is $P_s(path(b, f)|T1) = 0.316$ (note that it is sufficient to consider the graph restricted to nodes $b$, $e$, $d$ and $f$ when listing subprograms for this query), but for complex problems this could consume large amounts of time and memory. ProbLog therefore follows different strategies to obtain success probabilities, which we will briefly discuss next.

\subsection{Exact Inference} 
\label{sec:exact}
As iterating over possible subprograms as done in Equation~(\ref{eq:success}) is infeasible for most programs, ProbLog's exact inference instead employs a reduction to a propositional formula in disjunctive normal form (DNF). As stated earlier, probabilistic facts can be seen as RVs, implying that a proof can be represented as a conjunction of such facts. The set of all proofs can then be represented as a disjunction, producing a DNF formula. The success probability then corresponds to the probability of this formula being true. In our example we obtain the formula $(e(b,e) \wedge e(e,f)) \vee (e(b,d) \wedge e(d,f))$ where $e/2$ denotes edge. Each proof's probability is calculated as the product of the probabilities of its facts. Following the simple logic of conjunction and disjunction we could infer that the summation of all proofs' probabilities will produce the final result. However, this is only true under specific conditions, namely if each possible world permits at most one proof of the query. PRISM~\cite{SATO01} requires that programs respect these conditions, which means that proofs have to be \emph{mutually exclusive} (w.r.t.~occurrence in possible worlds). In our example, these conditions are not met:  we would obtain $0.34$, while the correct value is $0.316$. One way to deal with this problem is to consider the conjunctions in the DNF sequentially, and to replace each proof or conjunction $\alpha_i$ by its conjunction with the negation of all the proofs after it, that is, by $\alpha_i \wedge \bigwedge_{j\geq i}\neg\alpha_j$. In this way, each possible world permits at most one such extended proof. Note however that the resulting formula needs further manipulation to be transformed into a sum of products which can be used for easy calculation. For the previous example this will produce:
\begin{align*}
  P_s(path(b,f)|T) = & P((e(b,e) \wedge e(e,f)) \vee (e(b,d) \wedge e(d,f))|T)\\
                   = & P((e(b,e) \wedge e(e,f)) \wedge \neg(e(b,d) \wedge e(d,f))) + P(e(b,d) \wedge e(d,f))\\
                   = &0.8 \cdot 0.3 \cdot (1-(0.2 \cdot 0.5))+ 0.2 \cdot 0.5 = 0.316.
\end{align*}

Unfortunately this type of technique is feasible only for small formulae. This problem is known as the disjoint-sum-problem (as it is concerned with making the contributions of the different parts of the summation non-overlapping) and is \#P-complete~\cite{VALIANT79}. The ProbLog system deals with it using Reduced Ordered Binary Decision Diagrams (ROBDDs), which are graphical representations of a Boolean function over a set of variables, which significantly extends scalability of inference. We refer to this method as \emph{exact}.

\subsection{Approximate Inference: Program Sampling} 
\label{sec:program_sampling}
Furthermore, there exist different alternative inference methods from exact. One such approach that uses Monte Carlo methods, is to use the ProbLog program to generate large numbers of random subprograms and use those to estimate the probability. More specifically, such a method proceeds by repeating the following steps:
\begin{enumerate}
\item sample a logic (sub)program $L$ from the ProbLog program
\item search for a proof of the initially stated query $q$ in the sample $L\cup BK$
\item estimate the success probability as the fraction $P$ of samples which hold a proof of the query
\end{enumerate}
The implementation of this approach for ProbLog, as described in~\cite{KIMMIG08}, takes advantage of the independence of probabilistic facts to generate samples lazily while proving the query, that is, sampling and searching for proofs are interleaved. To assess the precision of the current estimate $P$, the width $\delta$ of the $95\%$ confidence interval is approximated as
\begin{equation}
 \delta = 2 \cdot \sqrt{\frac{P \cdot (1-P)}{N}}
\end{equation}
If the total number of samples $N$ is large enough the interval of confidence becomes smaller, and the certainty that the estimate is close to the true probability of the query increases. We refer to this method as \emph{program sampling}.

\subsection{Invoking Inference}
Probabilistic inference in ProbLog is invoked through \verb|problog_inference/3|. Given the inference method and the query one retrieves the probability that the query succeeds. When the given query is non-ground, \verb|problog_inference/3| computes as success probability the success probability of all the instances of the query without binding the variables. Later at Section~\ref{sec:problog_answers} we present a new inference method that returns the answers through backtracking by binding the non-ground variables.

\section{Why Probabilistic Meta-calls}
\label{sec:motivation}

Many real life applications use probabilistic inference to take decisions about a task. For a probabilistic logic system to fully support decision taking the nesting of its inference methods is required. Consider for example the problem of inferring the similarity between two words. While there are many approaches to tackle this problem, there is no best one. A reasonable approach is to infer the word similarity in different independent ways and then use a combination model. One could represent the synonym relation between words as a probabilistic graph and write a ProbLog program to infer the probability of two words having the same meaning. This technique does not perform well if spelling errors appear in the words. One could write another ProbLog program to find the probability of a spelling error. There are several ways to use these results in a probabilistic model. The final model that uses probabilistic inference during probabilistic inference, can be looked at as a higher order model.

The existing ProbLog implementation does not support nested inference. Moreover, once we start using nested inference we also want to determine at run-time the actual ProbLog query. We call the proposed extensions probabilistic meta-calls.

In Prolog, goals are nested all the time as the basic step of SLD resolution proofs a goal by proving the goals in the body of a unifying clause. Moreover, goals can be constructed as Prolog terms at run-time and then Prolog's support for meta-calls transforms the terms into executable goals.

The existing ProbLog system can compute a probabilistic query of the form \verb|problog_inference(Inference, Query, Result)| for a given $Query$, with a chosen $Inference$ method and returns the success probability at $Result$. Note that during this inference no nested calls to \verb|problog_inference/3| are allowed as they interfere with the bookkeeping of the use of probabilistic facts.

In this paper we propose to generalise the \verb|problog_inference/3| predicate, to allow nested inference and to support meta-call features. In addition of determining at run-time the inference method and the query, we also want to specify what kind of result we want: the success probability of the query, or a specific representation of the bookkeeping information on which the probability computation is based. The generalised ProbLog query will call \verb|problog_inference(| \verb|Inference, Query, ResultType, Result)| and these calls can be nested. We introduce the notion of a ProbLog engine that allow us to implement the general \verb|problog_inference/4| predicate.

\section{ProbLog Engine}
\label{sec:problog_engine}

Before describing the ProbLog engine, we want to point out that for the exact inference of Section~\ref{sec:exact}, ProbLog collects the probabilistic facts used in a success branch of the SLD tree of the query in a list (a so-called explanation or proof), and also collects all explanations as a DNF, which is typically represented by a trie. This trie is then transformed into a ROBDD in order to compute the correct success probability. On the other hand program sampling of Section~\ref{sec:program_sampling} samples a possible world which is kept in a list\footnote{Or an equivalent data structure like an array.} and counts the successful derivations.

Some shortcoming of the previous system are: the lack of intermixing different inference methods; that each inference method has its own SLD-resolution kernel; not easy to use alternative data structures; difficult to extend or modify current functionality. We identified the need for an abstract framework that provides a common SLD-resolution kernel that can be instantiated to realise different inference methods and/or different design options.

The ProbLog engine is an abstraction that allows dynamic modifications of Prolog's SLD resolution to uniformly implement the different bookkeeping needed by the different inference methods. By parametrisation of the ProbLog engine, one parametrises the SLD resolution.

The basic functionality of the parametrised ProbLog engine is the SLD based execution of the query together with the bookkeeping about the probabilistic facts that are used during this execution. By setting the parameters of the ProbLog engine, different instances are created that correspond to different inference methods and to different result types.

The difference with the existing implementations is the parametrised design, but also the organisation of the data-structures of the ProbLog engine. Each instance of the ProbLog engine has its own unique identifier, which is used when working with instance specific data. This instance-based organisation is necessary for the nested inference.

In the rest of this section we will explain what are relevant parameters of the ProbLog engine and how the nesting is supported.

\subsection{Parameters of the ProbLog Engine}
\label{sec:engine_parameters}
In order to define an instance of a ProbLog engine that implements an inference method, we specify two ``continuation'' predicates that deal with the construction of the explanations and the construction of the DNF. We also decide about the kind of data structures that are used to represent the explanations and the DNF. The instance of the ProbLog engine uses two ``registers'' to refer to the two data structures.

More specifically, we use two ``continuation'' predicates that are used during the SLD resolution to implement the adequate bookkeeping for the probabilistic facts, which are called annotated facts in this context: a fact annotated with a probability is a probabilistic fact. Both predicates are used to perform inference specific tasks and are different from inference to inference method.

The first dynamic ``continuation'' predicate is \verb|continuation_af/2| which is called every time a goal is proved by an annotated fact. The first argument is the unique identifier of the annotated fact and the second argument the annotation of the fact, typically its probability. The second dynamic ``continuation'' predicate is \verb|continuation_explanation/0| and is called every time the SLD resolution reaches a successful derivation. In addition to the two predicates, each instance of a ProbLog engine has two ``registers''. These registers refer to the instance specific data structures in which the information about the usage of annotated facts is collected. The first register (actually the referred data structure) is used by \verb|continuation_af/2| and the second by \verb|continuation_explanation/0|. These registers will be part of the state of the ProbLog engine that has to be saved and reset for nested inference.

In figure~\ref{fig:problog_engine} we present the parametrisations that implement exact and program sampling inference methods. For exact, the \verb|continuation_af/2| predicate is responsible for collecting the identifiers of the used annotated facts in a list, i.e. the $ID$ of the used annotated fact is added to the current explanation (referred to by the first register). The \verb|continuation_explanation/0| predicate is responsible for collecting the explanations in a trie: it adds the current completed explanation to the trie (referred to by the second register). The first register refers to the current (partial) explanation, and the second to the trie under construction.

For program sampling, the \verb|continuation_af/2| predicate is responsible for checking if the annotated fact or its negation is in the current sampled possible world, and if not to sample it and add it. The \verb|continuation_explanation/0| predicate does not need to do anything special. In this example we represent a partial possible world by a list but it could be represented by an array or another data structure. The first register refers to the current partial possible world. The equivalent bookkeeping for the second register would be the complete possible world, but because it is not required to sample the complete possible world we only need and keep a unique identifier which refers to the sample.

\begin{figure}[ht]
  \begin{verbatim}
  Continuation_af = (continuation_af(ID, _Probability):-
                       add_to_explanation(ID)),
  Continuation_explanation = (continuation_explanation:-
                                add_to_trie(completed)),
  problog_engine_init(exact,
                        continuations(Continuation_af,
                                      Continuation_explanation),
                        state(list, trie)).

  Continuation_af = [(continuation_af(ID, _Probability):-
                        in_possible_world(ID, Result),
                        !, call(Result)),
                     (continuation_af(ID, Probability):-
                        sample(Probability, Result),
                        add_possible_world(ID, Result),
                        call(Result))
                    ],
  Continuation_explanation = (continuation_explanation),
  problog_engine_init(program_sampling,
                        continuations(Continuation_af,
                                      Continuation_explanation),
                        state(list, identity)).  \end{verbatim}
\caption{Modifying the SLD resolution for exact, program sampling inference methods.}
\label{fig:problog_engine}
\end{figure}

\subsection{Nesting ProbLog engines}
The nesting of ProbLog engines, requires a suspension/resumption mechanism for which we use a stack. The active ProbLog engine is called the current engine.

When a new engine is initialised, it first pushes the current engine on the top of the stack and then becomes the active engine. When an engine has finished all its computations, it ends by popping the stack of engines.


When pushing an engine on the stack, we save the engine parameters: \\ \verb|continuation_af/2|, \verb|continuation_explanation/0|, the two registers, and the unique identifier of the engine. This information is sufficient to implement the nesting of engines using a stack discipline. Note that the information kept in the stack is related to the probabilistic bookkeeping. The interference of the nesting with the SLD resolution needs no special care.

\subsection{Calling the ProbLog Engine}
An instance of ProbLog engine is used to execute the \verb|problog_inference/2|, \verb|problog_inference/3|, and \verb|problog_inference/4| predicates. The four arguments of the predicate \verb|problog_inference/4| are $Inference$, $Goal$, $ResultType$, $Result$. The argument $Inference$ is used to define which ProbLog inference method is used and its possible values are $pure$, $exact$, $program\_sampling$, and  $current$, where $pure$ denotes that the engine behaves as pure Prolog and $current$ instructs the engine to use the same inference method as was being used. The argument $Goal$ denotes the goal that needs to be proved by the call. $ResultType$ denotes that we are interested in the success probability of the goal or in the explicit representation of the collected information and its possible values are $probability$ or $info$. Finally, the argument $Result$ is the returned probability or the detailed information that the engine returns.

We also use as syntactic sugar the predicates \verb|problog_inference/3| and \verb|problog_inference/2|.
\begin{verbatim}
problog_inference(Inference, Goal, Probability):-
  problog_inference(Inference, Goal, probability, Probabilistic).
problog_inference(Goal, Probability):-
  problog_inference(current, Goal, probability, Probabilistic).
\end{verbatim}

\section{Nested Inference}
\label{sec:meta-calls}
With our ProbLog engine based approach we can realise several interesting extensions as is described in the following subsections. Nested inference allows us to compute the success probability of a new child query that was possibly defined at the run-time, and use the success probability during the execution of the parent ProbLog query. The nested inference allows us to interleave any combination of different inference methods or inference tasks. Furthermore, returning the explicit information instead of the success probability can be used to formulate the counterpart of the \verb|\+/1| predicate of Prolog. Finally, we use our approach to support non-ground queries.

\subsection{Nested Inference Returning Success Probability}
Our approach allows us to perform nested inference. The nested inference computes the correct results as every call to \verb|problog_inference| suspends the previous ProbLog engine, starts a new one and uses the result when the previous one is resumed. In the example of figure~\ref{fig:inference_example}, ProbLog inference is used to decide which route will be taken.

\begin{figure}
  \begin{verbatim}
  ?- Input =..., problog_inference(exact, model(Input), Psucc).

  model(Input):-
    some_computation(Input, From),
    decide_route(From).

  decide_route(From):-
    problog_inference(path(a, From), P),
    (P < 0.3 ->
      path(From, f)
    ; P < 0. 6 ->
      path(From, g)
    ;
      path(From, h)
    ).  \end{verbatim}
  \caption{An example of inference within inference using the ProbLog program of figure~\ref{fig:problog_example}.}
  \label{fig:inference_example}
\end{figure}

\subsection{Nested Inference Returning Information \& ProbLog Negation}
The implementation of Negation as failure in Prolog uses a meta-call as shown in figure~\ref{fig:Negation}a. In ProbLog negation~\cite{Kimmig09srl} is a more complex task due to bookkeeping issues. In the existing implementation of ProbLog only probabilistic facts could be easily negated as their integration is very simple. Our approach allows to negate all probabilistic goals, namely goals that use in their proofs probabilistic facts.

Different inference methods require different implementations of negation. We illustrate the difference using exact and program sampling inference methods.

For exact inference, proving the negation of a probabilistic fact means to add to the explanation the complementary probabilistic fact\footnote{Probabilistic facts are represented as random variables, negating them results to the complementary random variable with probability $1-P$.}, as probabilistic facts are represented by Boolean variables, we simply mark them negated. This negation obviously does not alter the representation of the DNF formula.

The success probability of a probabilistic goal $Goal$ is computed from a representation of its corresponding DNF. In order for the negation of a probabilistic goal \verb|problog_not(goal)| to succeed, the negation of the corresponding DNF should hold or the corresponding CNF should hold.

Now consider the contribution of the subgoals to the final DNF: their corresponding annotated facts are scattered over the different explanations in the DNF. If we allow \verb|problog_not| in ProbLog, we need to do something special to incorporate the CNFs in a correct way. Moreover the calls to \verb|problog_not| can be nested. Our solution uses the suspend/resume mechanism to compute the DNF for the negated probabilistic goal. We also use a special data structure to represent the DNF: by collecting the explanations of the subgoals separately we will store them as is done for tabling~\cite{231065,mantadelis_et_al:LIPIcs:2010:2590} and in these nested tries we can indicate which parts need to be negated.

In the case of program sampling, things are very different. Instead of proof collection, we count in how many samples the query succeeds. In the process of constructing a possible world, we sample each probabilistic fact that we need to prove. At each sample, a probabilistic fact either succeeds or fails. Negating a probabilistic fact means inverting success with failure and failure with success. Similarly, a probabilistic goal succeeds or fails depending the possible world sampled, and its negation again means the invertion of success and failure. For that reasons one can use the negation as failure as defined in Prolog for ProbLog programs when doing program sampling.

\begin{figure}
  \begin{tabular}{ll}
\verb|'\+'(Goal):-|          &\verb|problog_not(exact, Goal):-|\\
\verb|  call(Goal), !, fail.|&\verb|  problog_inference(current,Goal,info,DNF),|\\
\verb|'\+'(_)|               &\verb|  continuation_af(not(DNF), _).|\\
                             &\verb|problog_not(program_sampling, Goal):-|\\
                             &\verb|  \+ Goal.|\\
\multicolumn{1}{c}{\textbf{(a)} Negation as Failure in Prolog} & \multicolumn{1}{c}{\textbf{(b)} ProbLog Negation}
  \end{tabular}
  \caption{Negation}
  \label{fig:Negation}
\end{figure}

\begin{figure}
\begin{verbatim}
  % This example encodes a coin toss.
  0.50::heads(_Number).
  toss(Number, heads) :- heads(Number).
  toss(Number, tails) :- problog_not(heads(Number)).
\end{verbatim}

\begin{verbatim}
  % This example encodes the following Sprinkler/Rain Bayesian network
  % from wikipedia.
\end{verbatim}
  \centering
  \includegraphics[scale=0.50]{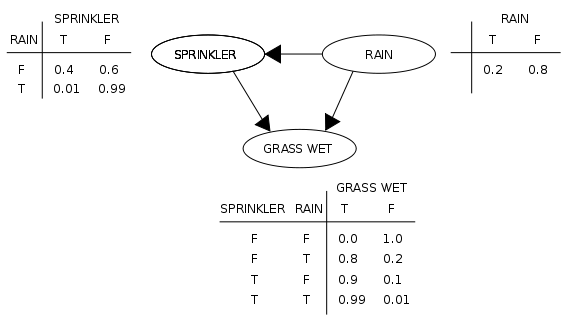}
\begin{verbatim}
  0.20::rain.
  0.01::sprinkler_on(rain).
  0.40::sprinkler_on(no_rain).
  0.80::grass_wet(rain).
  0.90::grass_wet(sprinkler).
  0.99::grass_wet(both).
  sprinkler :- rain, sprinkler_on(rain).
  sprinkler :- problog_not(rain), sprinkler_on(no_rain).
  grass_wet :- problog_not(sprinkler), rain, grass_wet(rain).
  grass_wet :- sprinkler, problog_not(rain), grass_wet(sprinkler).
  grass_wet :- sprinkler, rain, grass_wet(both).\end{verbatim}
  \caption{Example uses of ProbLog negation}
  \label{fig:Negation_examples}
\end{figure}

ProbLog negation has many uses in modelling. First of all it can be used to calculate the probability of a query not succeeding. Further on, it has been used to model annotated disjunctions which are needed for many probabilistic models such as a hidden Markov models, Bayesian networks and other. Some example uses are shown in Figure~\ref{fig:Negation_examples}.

\subsection{Nested Inference Returning Answers \& ProbLog Answers}
\label{sec:problog_answers}
Finding all the answers of a non-ground query in Prolog is done through backtracking. In ProbLog, we also need to calculate the success probability of the query having a particular answer.

We implemented this task by using Prolog to find the answers of the query. Once we have an answer, we have a grounding of the query and we can do probabilistic inference for this ground query. A simplification of the actual code implementing ProbLog answers is shown in Figure ~\ref{fig:problog_answers}. With this extension we can return answers to non-ground queries tupled with their success probability. We call this extension \emph{ProbLog answers}.

The \verb|call(Goal)| goal is using Prolog's backtracking mechanism to enumerate all possible answers by fully ignoring any probabilistic information related with the query. When an answer is found, \verb|problog_inference(Inference,Goal,P)| uses the appropriate inference method to calculate the probability of the answer.

This simplified code calls \verb|problog_inference/3| also when a particular answer occurs again. This inefficiency is solved easily by memoizing the calculated answers.

We need to add some functionality to the stack discipline. After dealing with one answer, we need to re-activate the parent engine\footnote{The outer engine that called ProbLog answers.} for continuing the previous goal, but when execution backtracks back to \verb|call(Goal)| we need to again activate the pure Prolog engine that returns us the answers. To solve this we implemented a special suspension mechanism which swaps the order of engines in the stack.

Finally, the last difficulty is that the different engines need separate garbage collectors which must be triggered when an engine is not needed anymore, thus we need a mechanism to tell us when the \verb|call(Goal)| is completed or the user commits to an answer\footnote{When for example the choice points are cut.}. This problem is solved with the help of YAP's \verb|setup_call_cleanup/3| built-in predicate.

ProbLog answers has many uses. Non-ground queries are needed to find which nodes are connected in probabilistic graphs and with which probability. One can use it in combination with other high order calls such as Prolog's \verb|findall/3|, \verb|forall/2|, etc. answering even more complex queries such as which node is more probable to be connected with a node in a probabilistic graph. See example in figure~\ref{fig:problog_answers_examples}.

\begin{figure}
\begin{verbatim}
problog_answers(Inference, Goal, P):-
  init_inference(pure_prolog_engine),
  call(Goal),
  problog_inference(Inference, Goal, P),
  (suspend_engine; (resume_engine, fail)).\end{verbatim}
\caption{Simplified ProbLog Answers}
\label{fig:problog_answers}
\end{figure}

\begin{figure}
\begin{verbatim}
connected_node(From, To, P):-
  problog_answers(path(From, To), P).

find_most_probable_node(From, MaxNodeTo, MaxP):-
  findall(To-P, problog_answers(path(From, To), P), Tuples),
  findmax(Tuples, MaxNodeTo, MaxP).\end{verbatim}
  \caption{Example uses of ProbLog answers}
  \label{fig:problog_answers_examples}
\end{figure}

\section{Experiments}
\label{sec:experiments}
Our experiments aim to measure the meta-call overheads. For the experiments we used a prototype\footnote{The prototype implementation is available at: \url{http://people.cs.kuleuven.be/~theofrastos.mantadelis/tools/metaproblog.tar.gz}} implementation of ProbLog that is implemented using the ProbLog engine approach. All the experiments are performed on an Intel Core 2 Duo CPU at 3.00GHz with 2GB of RAM memory running Ubuntu 8.04.2 Linux under a usual load using Yap 6.2.0~\cite{yap}.

To measure the overhead we used a typical ProbLog application namely an Alzheimer graph from~\cite{KIMMIG08} with a \verb|path/2| predicate that defines paths between nodes; we consider the pair of nodes \verb|'HGNC_582'|, \verb|'HGNC_983'| which is a query that has both many failing and succeeding derivations. We executed three different benchmarks, all of them nested with meta-calls exactly 10 times. The first benchmark is performing all the nesting first and then the goal is proved. The query for the first benchmark is of the form: $exact((exact((exact((...)),G1)),G1))$, where exact/2 is the abbreviation for \verb|problog_inference(exact,Goal,_P)|. This benchmark measures the overhead of the created engines.

The second benchmark has the goal before the nesting. In this way, an engine consumes resources before starting a nested engine. To avoid executing the nested call after each successfull derivation of the goal, \verb|path('HGNC_582',| \verb|'HGNC_983')|, we transform the goal into \verb|((path('HGNC_582','HGNC_983'),| \verb|fail);true)|\footnote{This will also reduce somewhat the work load but still retain it suitable for our experiment.}. The query for this benchmark is of the form: $exact((G2,exact((G2,$ $exact((...))))))$. This benchmark measures the the impact of previously proven goals to newer engines.

Our final benchmark is the combination of the two above. The query is of the form: $exact((G2,exact((G2,exact((...)),G1)),G1))$.

Executing the presented queries with no nested meta-calls we achieve the following execution times: $19069$, $12080$, $31191$ milliseconds respectively for the first, second and third query.

\begin{table}
\centering
\begin{tabular}{r|rr||r|rr}
\multicolumn{6}{l}{\bf{Query: exact((exact((exact((...)),G1)),G1))}}   \\\hline
Depth & Avg Time  & Deviation & Run & Avg. Time & Standard \\
  & (msec)  &   &   & (msec)  & Deviation \\\hline
1 & 19347.7 & 191.72  & 1 & 19447.2 & 125.89  \\
2 & 19281.3 & 120.29  & 2 & 19382.4 & 312.21  \\
3 & 19380.7 & 202.51  & 3 & 19287.2 & 216.96  \\
4 & 19279.2 & 114.64  & 4 & 19253.2 & 85.79 \\
5 & 19318.3 & 215.61  & 5 & 19231.9 & 144.35  \\
6 & 19240.1 & 155.73  & 6 & 19226.2 & 190.78  \\
7 & 19435.6 & 175.61  & 7 & 19561.5 & 103.54  \\
8 & 19268.2 & 99.21 & 8 & 19320.0 & 140.81  \\
9 & 19338.0 & 112.58  & 9 & 19285.8 & 89.11 \\
10  & 19361.3 & 340.75  & 10  & 19255.0 & 91.08 \\\hline
\multicolumn{6}{l}{\bf{Query: exact((G2,exact((G2,exact((...))))))}}   \\\hline
Depth & Avg Time  & Deviation & Run & Avg. Time & Standard \\
  & (msec)  &   &   & (msec)  & Deviation \\\hline
1 & 12202.5 & 124.22  & 1 & 12252.6 & 154.03  \\
2 & 12301.2 & 145.41  & 2 & 12269.5 & 73.19 \\
3 & 12269.2 & 130.31  & 3 & 12275.9 & 141.36  \\
4 & 12294.7 & 134.93  & 4 & 12240.8 & 108.59  \\
5 & 12256.9 & 163.27  & 5 & 12218.3 & 188.52  \\
6 & 12189.6 & 152.83  & 6 & 12272.3 & 176.31  \\
7 & 12169.9 & 140.61  & 7 & 12259.7 & 196.27  \\
8 & 12148.4 & 134.18  & 8 & 12234.1 & 141.92  \\
9 & 12312.4 & 119.22  & 9 & 12257.6 & 165.59  \\
10  & 12346.9 & 95.93 & 10  & 12210.9 & 76.06 \\\hline
\multicolumn{6}{l}{\bf{Query: exact((G2,exact((G2,exact((...)),G1)),G1))}}   \\\hline
Depth & Avg Time  & Deviation & Run & Avg. Time & Standard \\
  & (msec)  &   &   & (msec)  & Deviation \\\hline
1 & 31509.1 & 250.17  & 1 & 31494.3 & 300.57  \\
2 & 31749.2 & 362.68  & 2 & 31444.8 & 205.96  \\
3 & 31667.5 & 242.61  & 3 & 31529.5 & 138.27  \\
4 & 31416.1 & 121.02  & 4 & 31406.4 & 218.04  \\
5 & 31379.6 & 145.27  & 5 & 31453.1 & 363.23  \\
6 & 31331.7 & 270.83  & 6 & 31680.6 & 433.29  \\
7 & 31930.7 & 301.64  & 7 & 31562.7 & 121.24  \\
8 & 31466.0 & 322.14  & 8 & 31686.2 & 330.58  \\
9 & 31588.4 & 121.49  & 9 & 31482.0 & 203.66  \\
10  & 31331.8 & 108.03  & 10  & 31630.5 & 397.61  \\\hline
\end{tabular}
\caption{Experimental results.}
\label{tbl:results}
\end{table}

The results of our benchmarks are presented in Table~\ref{tbl:results}. The left hand side of the table presents the average times of 10 executions for each call at each nesting. The right hand side presents the average time any nested goal took at each execution. First thing we notice is that the all our averages are very close and that the standard deviation is low. From this observation we can safely claim that the depth of nesting does not impose any significant loss of time. On the other hand we do notice a very small overhead around $1\%$ going from no nested calls to any nested call.

\section{Conclusions, Related and Future Work}
\label{sec:conclusions}
In this paper we presented an approach for implementing probabilistic meta-calls using ProbLog engines. The underlying idea is to abstract the required information for the probabilistic bookkeeping in a parametrised ProbLog engine. By storing the engines in a stack, we achieve probabilistic meta-calls. We introduced the general probabilistic query \verb|problog_inference/4|, which allows us to perform nested inference and, further more, we presented how to implement \verb|problog_not/1| and \verb|problog_answers/3| with meta-calls. We also briefly illustrated the functionality of meta-calls in probabilistic modeling. To verify our approach, we performed some experiments and measured an overhead of approximately $1\%$.

ProbLog is closely related to other probabilistic logic systems such as PHA~\cite{POOLE93}, PRISM~\cite{SATO01}, PITA~\cite{RigSwi10-ICLP10-IC}, and ICL~\cite{POOLE95}. However, PRISM and PHA impose additional assumptions to simplify probability calculation, and the ICL implementation ailog2 does not scale to larger problems. ProbLog's implementation is targeted at overcoming these limitations. Unfortunately none of the existing probabilistic logic systems has support for probabilistic meta-calls. While PRISM has simpler bookkeeping (using support graphs) we believe that a similar approach like ours would be necessary to implement meta-calls, PITA is a specific tabled inference approach that very much resembles the tabling used by ProbLog~\cite{231065,mantadelis_et_al:LIPIcs:2010:2590}.

As future work, we are researching in methods for engine sharing. This approach aims at re-using evaluations among identical ProbLog engines imporving perfromance. Also, we are investigating the suitability of the ProbLog engine for other inference methods and we plan to implement them.

\section*{Acknowledgements}
We want to thank Paulo Moura for his advices on how to avoid nasty hacks in the implementation and the anonymous reviewers for their constructive comments. This research is supported by GOA/08/008 ``Probabilistic Logic Learning".

\bibliographystyle{splncs03}
\bibliography{CICLOPS2011}

\end{document}